Utilizing synthetic training data for the supervised classification of rat ultrasonic vocalizations.

*Running title*: *Synthetic-data for supervised classification of ultrasonic-vocalizations*

K. Jack Scott[+], Lucinda J. Speers[1], and David K. Bilkey[+]

[+] Department of Psychology, University of Otago, New Zealand.

[1] Grenoble Institut des Neurosciences, Inserm, France.

All correspondence to:
David K. Bilkey

David.Bilkey@otago.ac.nz

Department of Psychology

William James Building, 275 Leith Walk

University of Otago, Dunedin 9016, New Zealand.

ORC ID:
K. Jack Scott: 0000-0002-8186-8578

Lucinda J. Speers: 0000-0001-8298-3781

David K. Bilkey: 0000-0002-1256-4888

Supported by funding from the University of Otago.

The following article has been accepted by the Journal of Acoustical Society of America. After it is published, it will be found at

https://doi.org/10.1121/10.0024340

K. J. Scott *et al.,* J. Acoust. Soc. Am. 1 January 2024; 155 (1): 306-314

## Abstract

Murine rodents generate ultrasonic vocalizations (USVs) with frequencies that extend to around 120kHz. These calls are important in social behaviour, and so their analysis can provide insights into the function of vocal communication, and its dysfunction. The manual identification of USVs, and subsequent classification into different subcategories is time consuming. Although machine learning approaches for identification and classification can lead to enormous efficiency gains, the time and effort required to generate training data can be high, and the accuracy of current approaches can be problematic. Here we compare the detection and classification performance of a trained human against two convolutional neural networks (CNNs), DeepSqueak and VocalMat, on audio containing rat USVs. Furthermore, we test the effect of inserting synthetic USVs into the training data of the VocalMat CNN as a means of reducing the workload associated with generating a training set. Our results indicate that VocalMat outperformed the DeepSqueak CNN on measures of call identification, and classification. Additionally, we found that the augmentation of training data with synthetic images resulted in a further improvement in accuracy, such that it was sufficiently close to human performance to allow for the use of this software in laboratory conditions.

**Acknowledgements**

We would like to thank animal technician Sophie French for assistance with animal care.

**JASA acoustic PACS code:** 43.60.Np

## I. Introduction

Over the past two decades researchers have become increasingly interested in the ultrasonic vocalizations (USVs) of rats and mice [7,21]. The social functions of these calls are, at present, not fully understood and so examining their detailed structure and the context in which they are produced will underlie progress in this area. Furthermore, recent research suggests that rat USVs may have some features that align with aspects of human vocal communication, allowing for investigations of some of the underlying mechanisms [26,29]. For example, USVs display a sequential structure that may depend on some of the same mechanisms that are affected in neurodevelopmental disorders such as schizophrenia and autism spectrum disorder [35], and in this context, USV structure may serve as a measure against which novel treatments for these disorders can be tested [21,32]. Furthermore, as USVs are associated with affective state, they are being used to explore emotional dysregulation in models of human disorder[34] However, in order to utilize rat USVs in these ways, it is important that researchers have access to appropriately accurate and rapid classification systems.

In adult rats, there are two broad categories of USVs. First, there is a category of 'aversive' USVs that are often labelled as being of the 22-kHz type. USVs of this type actually occur within a frequency band of 18-30kHz, are long in duration (between 500ms – 2.5s), are flat in frequency, and tend to occur in a series or 'bout' of calls [4,21]. These 22-kHz USVs tend to occur during stressful situations such as when rats sense the presence of natural predators, or during fear conditioning in a shock chamber [23,43]. Thus, it is hypothesised that 22-KhZ calls serve an alarm or distress function [6]. By contrast, a second group of '50-kHz' calls occur over a much wider frequency band (30 – 90-kHz), are heterogenous in frequency contour, and are comparatively short in duration, with most calls lasting for between 5 and 90ms. These calls tend to occur in appetitive or rewarding situations [6,21].

There have been several classification schemes proposed for rat 50-kHz calls. One of the simplest defines two categories; flat and frequency modulated calls. A flat call is defined as a signal which does not vary substantially in frequency (kHz) whereas a modulated call would rise or fall in frequency [9,38]. Burgdorf and colleagues (2008) and later Brudzynski (2015) favoured a classification scheme determined by behavioural correlations, which includes flat calls, frequency modulated step-based calls, and frequency modulated trill-based calls[5,8]. In contrast, Wright et al. (2010) classified 50-kHz calls according to the structure and frequency contour of calls, and separated 14 distinct sub-types [41]. Although the behavioural significance of many of these call types in unknown, certain calls classified using Wright and colleagues' schema have been linked with specific behaviours. For example, in one study, researchers found that Trill calls, upward ramp and short calls predominated the call inventories of juvenile rats engaged in play, and signalled the onset of specific aspects of play [19]. Additionally, Burke and colleagues (2017) found that several Wright style classifiers were highly correlated with specific behaviours during anticipation of play [10]. Furthermore, these researchers found that commonly expressed calls classified using Wright style classifiers could be clustered into 4 groups according to their behavioural correlation [10]. Fifty kHz calls may also be elicited in non-social contexts, for example following the administration of psychoactive drugs. Psychostimulants such as amphetamine have been found to modulate the expression of trill and flat calls, with trill calls increasing and flat calls decreasing following administration of amphetamine, suggesting that trill based calls may indicate a positive affective state[41]. However, in a separate study, amphetamine and methylphenidate increased the expression of all 50-kHz call subtypes[33]. Interestingly, Wright and colleagues (2012) reported that the administration of morphine at rewarding and hyperlocomotion-inducing doses to rats did not induce an increase in trill based calls, indicating that the expression of these calls may not be tied directly to positive affective states in all instances [40].

The process of manually identifying and classifying call types is both time consuming and resource intensive [13]. Thus, in recent years researchers have turned to neural networks to expedite these processes. Convolutional neural networks (CNNs) such as DeepSqueak [13], have substantially cut down on analysis time. Previously, in our lab we have employed this software for automated detection and supervised classification of rat USVs [32]. However, recent research has drawn attention to potential deficiencies in DeepSqueak, both for identification and supervised classification of USVs, relative to more recent software developed for mice USVs [16,17,37]. VocalMat (VM; Fonseca et al., 2021) is one such software package, that has previously been compared to DeepSqueak (DS) for the detection of mice USVs [16]. The researchers found that VM outperformed DS on several measures of detection, with an experienced observer's ratings used as ground truth. These findings suggest VocalMat might hold advantages for detection of USVs over DeepSqueak. Additionally, Fonseca and colleagues reported that VM performed better than DS on mouse call classification, using a classification scheme that was functionally similar to Wright style classifiers in both classification method, and diversity of categories [1,16]. Here we test the performance of VM compared to DS for the detection and classification of rat USVs.

The process of manually classifying USVs in order to build a suitable set of data with which to train classification networks is time consuming [3]. One method that has been previously used in the wider neural network field to facilitate the development of a training dataset is to augment the curated data with automatically generated synthetic analogues. These synthetic data have the same essential characteristics as the target, but have been altered within the natural variation of the source. This process has been used in diverse applications, such as 3D body and object pose estimation, face and text recognition, and traffic recognition [14,20,28]. Here we assess whether augmenting a training set of natural USVs with synthetic data improves the accuracy of the VM classification network.

## II. Methods

Subjects

Sixty-eight adult male Sprague-Dawley rats were recorded to produce the dataset used for the present experiment. Thirty-two animals originated from 11 litters where the mother had been treated with poly I:C on day 15 of the pup's gestation to model maternal immune activation (MIA), while 36 animals came from 10 saline-vehicle litters. The MIA manipulation was for the purpose of other experiments not described here. Full details of the procedure are available from previous research in our laboratory [22,39]. In our experience MIA produces minimal changes in USVs, apart from a prolongation of 50 kHz call duration that is within the upper bounds of control calls[32]. For these reasons calls from MIA and control animals were treated as one group. Offspring were housed in litter-mate groups until reaching the age of maturity (3-months) with ad-libitum access to food and water. Upon reaching maturity, these male offspring were separated into housing cages with a paired-condition litter-mate, and maintained on food deprivation of no less than 85% of the animal's recorded free-feeding weight, whilst having access to water ad-libitum. Animals were maintained on food deprivation in order to simulate naturalistic satiation conditions. These animals were housed in their home cages in a room which was temperature controlled, with lighting following a 12-hour light/dark cycle. All animals were recorded during the light phase of this lighting cycle.

Experimentation was approved by the Otago University Animal Ethics Committee in accordance with the NZ Animal Welfare Act 1999.

Apparatus and materials

Rat USVs were recorded from animals placed in a rectangular box with two chambers. A clear plexiglass partition perforated with approximately fifty 10mm holes separated the two chambers. The dimensions of each chamber were 600mm long, 240mm wide, and 600mm high. Ultrasonic vocalizations were recorded with a pair of UltraMic 250k USB microphones (Dodotronic, Italy), which were affixed to the rear chamber walls at a 45-degree angle to the floor. These microphones recorded audio (sampled at 250 kHz), to laptops running Audacity for Windows (3.0.0). USVs in the audio files were classified by two different convolutional neural networks; DeepSqueak (v3.0; Coffey et al., 2019) and VocalMat (v1.0; Fonseca et al., 2021), both sourced from repositories on Github. Both DeepSqueak and VocalMat operated on MATLAB (2021A, Mathworks) running off Windows 10. A modified version of the Wright et al. (2010) classification system was employed as per our previous research [32], which was a default classification network available in DeepSqueak (v2.6x). Calls were categorized into the following categories: Complex, Complex trill, Downward ramp, Flat, Inverted-U, Short, Split, Step down, Step up, Trill, and Upward ramp.

These call categories represent a slightly simplified version of the classification scheme originally proposed by Wright and colleagues (2010). Here, multi-step calls have been combined with split call types, now defined as a sudden drop or rise of frequency that returns to the starting range. Trills with jumps from the original Wright et al. (2010) schema have been conflated into the trill sub-type and both composite and flat-trill combination call types from Wright et al. (2010)'s schema have been merged within the new category complex trill[13], although it bears noting that we have not observed composite calls of the length and complexity as displayed in Wright et al. (2010)'s exemplars. Whilst not exhaustive in the categorization of all USV emissions, this categorization captures the overwhelming majority of adult Sprague-Dawley calls detected across our recordings within the 50-kHz band. Additionally, the classification scheme represents a compromise that maximizes classification accuracy, whilst preserving the essential syntactic structure of rat vocalizations (Figure 1). Data wrangling and processing was conducted in MATLAB (R2021A) and Excel (Microsoft, 2019). GraphPad Prism (9.0) was utilized for data-analysis and statistical testing.

**Fig. 1**

*Simplified classification schema for Rat USVs within the 50-kHz band*

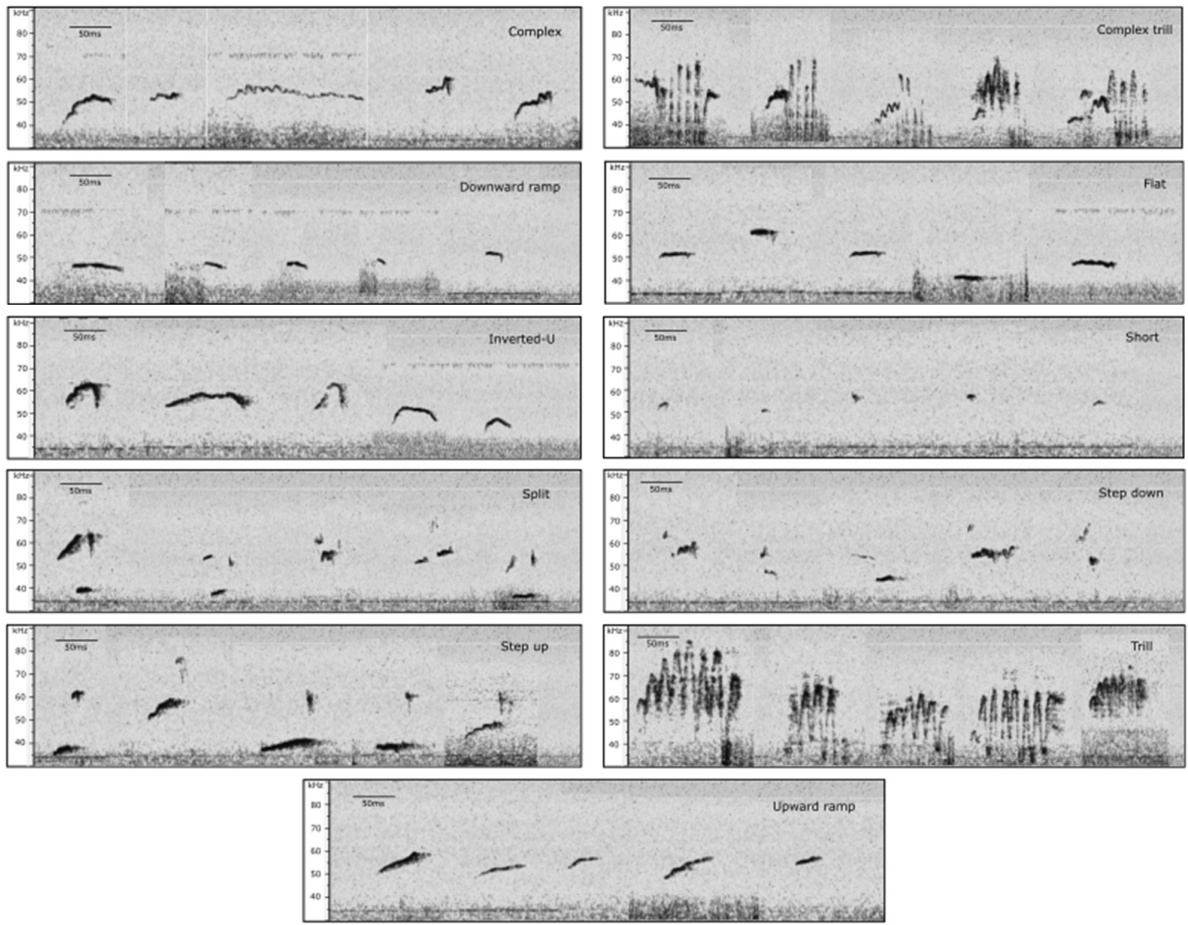

Procedure

      For each recording either one pair of matched MIA, or one pair of matched control animals were selected from different home cages. An intervening cage selection criterion was employed to ensure equidistance of animals selected for testing, and to account for familiarity as an extraneous variable of USV emission [32]. Animals were habituated to the recording chamber a day before testing for a period of ten minutes, followed by a second

habituation period of five minutes prior to recording. Habituation was conducted in this manner as previous research indicates animal stress levels are a variable affecting both USV emission rate and type of calls emitted [11]. Animals were placed one in each chamber, and recorded for periods that ranged between 10 and 13 minutes before being returned to home cages. These recordings were exported in .wav format and utilized for analysis.

*Detection of USVs*

A comparison of DeepSqueak (v.3.0; Marx et al., 2021) and VocalMat (v1.0; Fonseca et al., 2021) was initially conducted to determine how effectively they discriminated USV calls from noise (prior to classification). Two different ten-minute recordings were utilized in this analysis, with one being a prototypically noisy recording, and the other being a low noise recording (refer to supplementary material, S1). These recordings were processed for USV detection by DeepSqueak and VocalMat, and the performance compared to that of a trained experimenter's judgment (JS). This allowed us to determine whether the detected signals were true USVs (hits) or false alarms, and whether misses had been made. Data was analysed using signal detection procedures [36].

*Classification of USVs*

DeepSqueak and VocalMat were initially trained for classification on the same data set. This dataset comprised spectrograms of calls and was produced by loading a file with detected USVs, and then manually labelling calls in DeepSqueak's spectrogram navigator

(see supplementary information S5). We selected recordings that had over 250 prototypical 50-kHz calls with diverse spectrotemporal structure. Approximately 10 exemplars per call type, per recording were selected. Seven such training files were created, which produced a dataset of 580 calls. These calls were then exported as training images and then rechecked to ensure all calls were aligned in the centre of the spectrogram image. Once this procedure was completed, 498 calls were retained for network training and validation. MATLAB code was written which batch converted images produced by DeepSqueak from grayscale to RGB, and then resized the images for VocalMat format. During training, a subset of images were labelled and presented to the network. This was followed by a validation phase where unlabelled images were presented to the network. Validation accuracy (%) was measured by comparing validation performance against the ground-truth labelling. Five naïve DeepSqueak classifier networks and five naïve VocalMat classifier networks were trained using this image set, and performance on validation compared between DeepSqueak and VocalMat networks.

*Classification with iterative training in VocalMat, and Synthetic images*

To further explore VocalMat performance, new call images were added to the training set at regular intervals while network training continued. We then tested whether augmenting this dataset with synthetic images would improve call categorisation performance. To these ends, MATLAB code was written that randomly morphed selected seed images to produce synthetic offspring (supplementary material, S4) using the morphimage routine [27]. Each morphed image was then quickly reviewed and retained on the basis of it being both substantially different from the source image, but also a viable replicant of a natural call of the source category. With the introduction of synthetic calls, the dataset was initially increased from 401 (trial 43; Figure 4) to 802 calls (401:401; natural:synthetic; trial 44). In a second stage, an additional 390 synthetic calls, and an additional 14 natural calls (the seed images) were added (trial 48; Figure 4), producing a 1,206 call dataset (415:791; natural:synthetic). This second augmentation was conducted to prevent network overtraining, and to assess whether further marginal gains in accuracy would be evidenced [2].

Once training had taken place, generalization tests were conducted to test the accuracy of trained networks when categorizing novel USV images. Classification was based on the modified Wright et al. (2010) categories, and the proportion of correct classifications were determined by comparing the automated classification with that of manual classification using skilled human judgement as ground truth. Five generalization tests were conducted following the introduction of synthetic calls to the training dataset. These tests were conducted on five recordings that had not been used for training and which contained a number of challenging and ambiguous calls. To determine whether synthetic call training had altered performance, and to assess over-training, we conducted a pre-post comparison where we compared the performance of a network trained immediately prior to the introduction of synthetic calls (trial 43), to networks trained following the introduction of synthetics (trials 44, 48; Figure 4).

Data analysis

*Comparing detection between DeepSqueak and VocalMat*

To assess how effective DeepSqueak and VocalMat were at detecting USVs, we conducted a signal detection analysis of detection of calls recorded in both a low-noise and high-noise environment. Signal detection theory is useful in this regard, as it has several measures for determining how well an agent (be it a human or a neural network) is able to discriminate between a true signal and noise [36]. The hit rate and false alarm rate were calculated. The hit rate is the number of correct detections divided by the total number of detections, whereas the false alarm rate is the number of detections where no true signal is present divided by the total number of detections. The hit rate and false alarm rate were then z-transformed and the d prime score was calculated (z-transform of hit rate – z-transform of false alarm rate). The d prime score is a summary measure of discrimination based on a comparison of hits and false alarms [18].

*Generalization tests assessing accuracy of VocalMat classification networks trained on natural relative to combined natural and synthetic data*

VocalMat produced data output for each detected call, including the classification of call-type. Each detected USV was then marked by a trained human observer as to whether it was judged to be a correct or incorrect classification. The proportion of correct classifications was calculated by totalling the number of correct classifications and dividing this number by the total number of classified USVs. The proportions of correct classifications were then analysed across the five generalization tests. Means and descriptive statistics were calculated in GraphPad Prism. A two tailed independent samples t-test was used to compare the accuracy on generalization of a network trained on natural data alone to those networks trained on a combined natural/synthetic training set across the five novel recordings.

## III. Results

USV Detection

The high-noise and low-noise USV recordings were run through both DeepSqueak and Vocal Mat networks. The resultant USV detection performance is summarized in Table 1 and Table 2.

**Table 1.**

*Inventory of calls for DeepSqueak (DS) and VocalMat (VM) across high and low noise recordings*

|  | Detections | Hits (True USV) | False Alarms | Misses |
| --- | --- | --- | --- | --- |
| DeepSqueak High | 1470 | 78 | 1392 | 33 |
| VocalMat High | 139 | 124 | 15 | 6 |
| DeepSqueak low | 201 | 175 | 26 | 13 |
| VocalMat low | 138 | 133 | 5 | 6 |

**Table 2.**

*Signal detection performance for DeepSqueak and VocalMat across high and low noise recordings*

|  | Hit rate | False Alarm rate | Miss rate |
|---|---|---|---|
| DeepSqueak High | 0.053 | 0.947 | 0.423 |
| VocalMat High | 0.892 | 0.108 | 0.121 |
| DeepSqueak low | 0.871 | 0.129 | 0.074 |
| *VocalMat low* | 0.957 | 0.036 | 0.037 |

As can be seen in Table 2, VocalMat had substantial performance advantages over DeepSqueak, and this advantage was particularly pronounced in the high-noise recording. To summarize the relative discrimination performance of DeepSqueak and VocalMat, a d prime score was computed across the two recordings. For the high noise recording VocalMat ($d' = 2.475$) substantially outperformed DeepSqueak ($d' = -3.59$). For the low noise recording, VocalMat ($d' = 3.511$) again outperformed DeepSqueak ($d' = 2.26$), however at a less dramatic difference (Figure 2). The negative d prime score for DeepSqueak in the high-noise condition indicated that the network was encountering substantial issues discriminating between noise and true USVs.

**Fig. 2**

*Relative detection performance of DeepSqueak, VocalMat (d prime)*

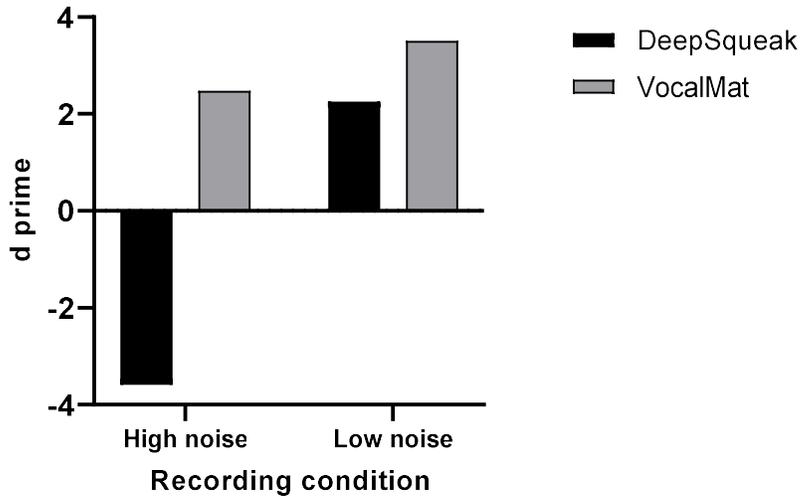

USV Classification

A total of 498 USVs, each manually classified by a skilled observer, were available in the seven recordings used for network training. The mean (SEM) number of USVs per recording was 71 calls (± 5.4). Across the five DeepSqueak classification networks trained on this dataset, the mean classification accuracy was 32.73% during the validation phase. VocalMat networks trained on the same dataset obtained a mean accuracy of 56.11%. An unpaired t-test revealed that this difference was highly statistically significant ($t(6.83)=8.403$; $p<.0001$).

**Fig. 3**

*A comparison of naïve trained network classification validation accuracy*

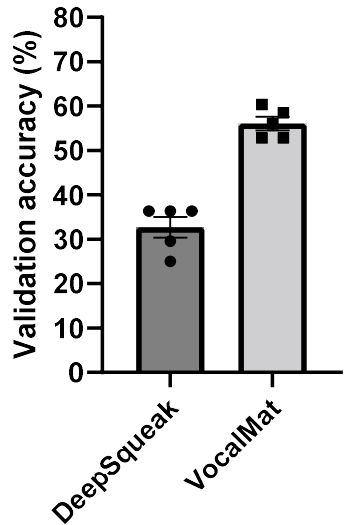

Supervised classification in VocalMat

We next explored the effect of extending training on the classification performance of VocalMat, utilizing training images with low background noise levels. With further training (trials 5 to 42), including intermittent expansion of the training set, training accuracy improved to around 95%. This improvement was mirrored in improvements in the validation procedure accuracy (peaking at 87.8%), where a subset of the data that the network has not been specifically trained on is tested (Figure 4).

**Fig. 4**

*Progression of accuracy (correctly classified USVs) on network training and validation for trained VocalMat classification networks*

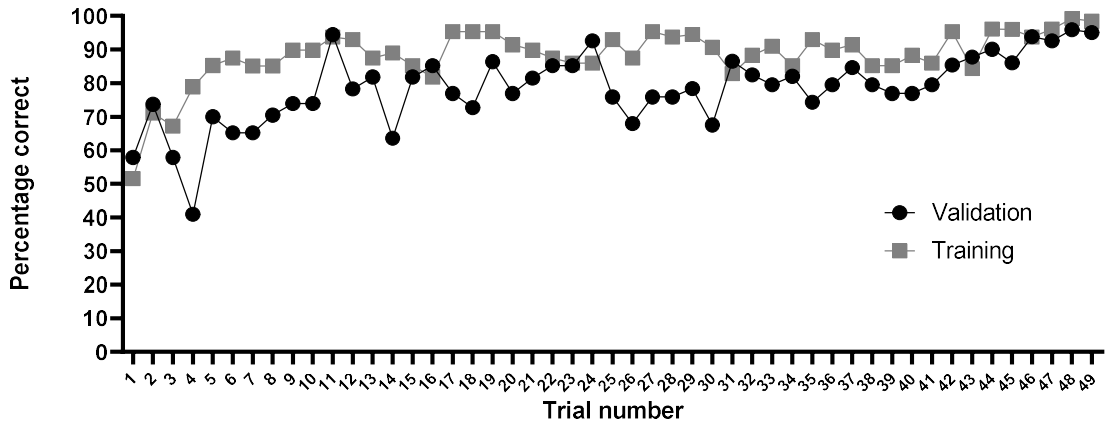

VocalMat classification performance after training on a combined synthetic/natural dataset

Network training requires exposure to exemplar USV calls which require a considerable amount of time and effort to manually classify. In order to speed up this process we generated 791 synthetic calls and captured 14 additional natural calls, and inserted them randomly into the initial pool of 401 natural calls, at two training trials (44, 48; Fig. 4). A series of five generalization tests were conducted by presenting a series of calls that the network had not previously been exposed to. The performance of the network trained immediately prior to the introduction of the synthetic calls (PRE) was compared to that of the network trained after the introduction of synthetic calls (POST). Across the five test recordings, the mean (SEM) correct categorizations for the PRE-network was 0.713 (± 0.01), whereas for the POST-networks the mean (SEM) was 0.826 (± 0.01). Thus, adding synthetic calls to the training set increased performance by a mean of 11%. A two-tailed paired t-test comparing proportion correct (%) between PRE and POST networks revealed that this improvement in correct classifications was statistically significant (t(7.52)=6.955, p=.0002). These findings are summarized in Figure 5 below.

**Fig. 5**

*Classification performance of networks trained on natural data alone, with those trained on a combined natural and synthetic dataset, across five novel recordings*

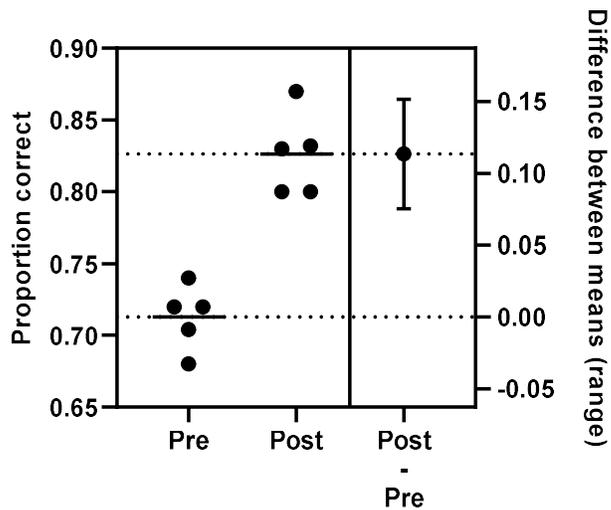

IV. Discussion

The primary aim of this research was to modify the CNN VocalMat [16] to detect and categorise rat USVs according to a classification system modified from Wright et al. (2010), and to compare the performance of this network with another CNN, DeepSqueak v3 [25]. Additionally, we sought to assess whether the introduction of synthetic USV calls into the VocalMat training set would improve correct classifications of novel data. We successfully adapted VocalMat for rat USVs and found that VocalMat outperforms DeepSqueak on call detection, as measured using signal detection procedures. We then tested supervised classification on a small dataset consisting of 497 USVs, finding that VocalMat networks significantly outperformed DeepSqueak classification networks. Finally, the addition of synthetic calls to the training set significantly improved performance of VocalMat classification networks on generalization accuracy in novel recordings.

Our results are consistent with a comparison previously conducted by Fonseca and colleagues (2021) utilizing USVs recorded from mice, where the authors found that VocalMat performed better than DeepSqueak on USV detection, although these researchers did not compare the two networks on classification. VocalMat's superior detection of USVs could be attributable to differences in workflow between VocalMat and DeepSqueak, with VocalMat employing image processing techniques and morphological operations to improve spectrogram images to better discriminate USVs from background noise. Additionally, while both DeepSqueak and VocalMat utilize a CNN for noise detection, VocalMat also includes an additional step in the workflow, using a local median filter to estimate the lowest anticipated contrast between a USV and background noise for a given recording before passing the spectrogram to the CNN for noise detection [16]. However, as we only examined performance on a relatively small number of calls, we cannot rule out the potential for variability in USV detection in different recordings.

We also found that VocalMat outperforms DeepSqueak in USV classification of rat USVs. We trained five naïve classification networks in both VocalMat and Deepsqueak on the same dataset of 497 USVs. VocalMat trained classification networks had a mean accuracy for correct classification in the validation phase that was 23.4% higher than DeepSqueak. This was a consistent effect across each network tested, although a caveat is that a training set of ~500 calls is relatively small for training classification networks. It is probable that training on a larger dataset would have yielded better results for both networks.  To illustrate, the original DeepSqueak (v1) classification networks were trained on USV inventories of approximately 40,000 calls [13]. However, public discussion with the developers revealed that even at such high call inventories for training, DeepSqueak trained classification networks did not perform above 70% on validation. It has been suggested that the problem may lie with the complexity of Wright et al. (2010) style classifiers, and a simpler classification system (i.e. with less categories) may yield better results [12]. Such classification systems have been proposed in the literature, for instance Brudzynski (2015) favours a simpler classification of three 50-kHz types according to behavioural correlation, while Wöhr and colleagues prefer a simple binary between flat and frequency modulated calls at 50-kHz [5,38]. Importantly, the accuracy of supervised classification is dependent upon sufficiently differentiated categories, which may underlie the concerns some researchers have in regards to Wright style classifiers[12]. Our findings that classification performance using a modified version of Wright et al. (2010) classifiers can be quite high may help address these concerns.

We iteratively trained a series of VocalMat classification networks, forking the prior-network for training, and adding new USVs to the training set at regular intervals. Early on in this procedure it became clear that relatively high accuracy on validation was indeed possible using our modified Wright-style classifiers [32], with accuracy on validation reaching between 80 – 90% in many trials. However, generalization to new recordings, while approximately similar to the validation reported by Coffey and colleagues, still fell short of what we considered to be an acceptable criterion for accurate classification. Additionally, network performance (validation accuracy) had remained relatively stable across several new iterations of classification networks trained only on the natural dataset.

We found, however, that accuracy on both training-set validation, and classification of novel recordings was significantly improved with the introduction of synthetic calls to the training set. To our knowledge, no prior USV classification research has attempted to train a classification network using a training set that includes synthetic USV spectrogram images. However, many researchers are successfully using synthetic data sets to accurately train other types of classification networks. For example, synthetic data has been used to train CNNs in diverse fields [14,15,42]. In one study, Dewi and colleagues report a similar performance improvement to the findings of the present study, with a CNN trained to recognise traffic signs improving from 69.7% accuracy on generalization to an average of 86.7% when synthetic images were introduced [14]. Given the labour-intensive nature of expanding the inventory of training sets for supervised classification [3], we show that the augmentation of training sets with morphed spectrogram images is an efficient and effective way to improve network accuracy for supervised classification of rat USVs.

With the inclusion of synthetic calls into the training set VocalMat reached a mean accuracy on generalization of 82.6%. This performance was achieved with a comparatively small training set relative to that used previously for mice USVs [16], but achieved a similar level of accuracy. Of note however, the recordings used for our generalization tests did not generally feature a strong signal, having been recorded at a lower gain (see supplementary material, S5), which may render classification more difficult in certain circumstances, but may also limit misclassification due to the absence of noise. In the present research we did not investigate whether CNNs find classification more difficult under low signal strength/low noise conditions or high signal strength/high noise conditions. Exploration of the effects of high and low noise recordings on VocalMat training and validation would be useful in future studies. Additionally, while the generalization performance we observed was similar to that reported for mice USVs in Fonseca et al. (2021), it remains an open question as to whether using a larger training dataset would yield substantial additional gains in accuracy. It may be beneficial for future research to test this using a variety of recordings with varying signal to noise ratios.

The present research describes the successful adaptation of VocalMat for analysing recordings of rat USVs. When compared to another widely used identification and classification network, we found that VocalMat offered better performance, following training with a small dataset. While a performance comparison might have produced different results with more extended training on a larger data set, the ability to use small training sets is an important time-saver. The addition of synthetic calls to the training set led to further improvement in VocalMat classification, indicating that this is a relatively easy way to improve performance. By providing accuracy, consistency and an ease-of-use process, CNNs open up new opportunities for detailed investigations of the role of USVs in communication [13], and research into the underlying structure and biological mechanisms [32]. For instance, recent research suggests that animal calls may have features that align with fundamental aspects of human language, allowing for investigations of the biological underpinnings of human vocal communication [29]. USVs also hold promise as a behavioural assay for examining models of neurodevelopmental disorders in humans such as schizophrenia and autism spectrum disorder as they display sequential structure that may depend on some of the same neural mechanisms that are affected in these disorders [35]. More broadly, interesting questions arise in the field of animal communication. While both human observers and CNNs can characterise calls into groupings such as those devised by Wright and colleagues [16,41], it remains to be determined whether these calls are interpreted as distinct by the animals themselves. While researchers are making headway with this question in mice [30], additional research in other animals may provide important insight for the field.

Declarations

The authors have no relevant financial or propriety interests in any material discussed in this article.

**Data availability**

The datasets generated during and/or analysed during the current study are available from the corresponding author on reasonable request.

Our adaptation of VocalMat for rats is forked from the original (https://github.com/ahof1704/VocalMat) released at https://github.com/scoki211/VocalMat4Rats

**Supplementary Material**

See supplementary material at https://pubs.aip.org/jasa/article-supplement/3023024/zip/306_1_10.0024340.suppl_material/ for exemplars of recording noise, classification scheme rationale, DeepSqueak manual labelling, synthetic call morphing exemplar, and signal to noise exemplar for recordings used for generalization tests.